\documentclass[pra,twocolumn,noshowkeys,preprintnumbers,amsmath,%
amssymb,showpacs,aps,10pt,superscriptaddress]{revtex4}
\usepackage{graphicx,color}

\usepackage{amsmath,enumerate}
\usepackage{bbm,amsfonts}

\newcommand{\be}{\begin{equation}}
\newcommand{\ee}{\end{equation}}
\newcommand{\bea}{\begin{eqnarray}}
\newcommand{\eea}{\end{eqnarray}}

\newcommand{\tr}{\mathrm{tr}}

\newcommand{\ket}[1]{\left|#1\right\rangle}

\newcommand{\hideit}[1]{}

\begin{document}

\title{Entanglement spectrum and boundary theories with projected entangled-pair states}

\author{J.~Ignacio \surname{Cirac}}
\affiliation{Max-Planck-Institut f{\"{u}}r Quantenoptik,
Hans-Kopfermann-Str.\ 1, D-85748 Garching, Germany}
\author{Didier Poilblanc}
\affiliation{Laboratoire de Physique Th\'eorique, C.N.R.S. and Universit\'e de Toulouse, 31062 Toulouse, France}
\author{Norbert Schuch}
\affiliation{Institute for Quantum Information, California Institute of
Technology, MC 305-16, Pasadena CA 91125, U.S.A.}
\author{Frank Verstraete}
\affiliation{Vienna Center for Quantum Technologies, Faculty of Physics, University of Vienna, 1090 Wien,  Austria}

\begin{abstract}
In many physical scenarios, close relations between the bulk properties of
quantum systems and theories associated to their boundaries have been
observed.  In this work, we provide an exact duality mapping between the
bulk of a quantum spin system and its boundary using Projected Entangled
Pair States (PEPS). This duality associates to every region a Hamiltonian
on its boundary, in such a way that the entanglement spectrum of the bulk
corresponds to the excitation spectrum of the boundary Hamiltonian. We study
various specific models, like a deformed AKLT \cite{AKLT1}, an Ising-type \cite{verstraetewolf06}, and Kitaev's toric code \cite{kitaev:toriccode}, both in finite ladders and infinite square lattices. In the latter case, some of those models display quantum phase transitions. We find that a gapped bulk phase with local order corresponds to a boundary Hamiltonian with local interactions, whereas critical behavior in the bulk is reflected on a diverging interaction length of the boundary Hamiltonian. Furthermore, topologically ordered states yield non-local Hamiltonians. As our duality also associates a
boundary operator to any operator in the bulk, it in fact provides a
full holographic framework for the study of quantum many-body systems via
their boundary.
\end{abstract}

\maketitle

\section{Introduction}

It has long been speculated that the boundary plays a very significant role in establishing the physical properties of a quantum field theory. This idea has been very fruitful in clarifying the physics of the fractional quantum Hall effect, and is also the origin of the holographic principle in black hole physics. An explicit manifestation of this fact is the so-called area law. The area law states that for ground (thermal) states of lattice systems with short-range interactions, the entropy (quantum mutual information) of the reduced density operator $\rho_A$, corresponding to a region $A$, is proportional to the surface of that region, rather than to the volume, at least for gapped systems \cite{Srendnicki,Wilczek2,Hastings,WVHC}. Criticality may reflect itself by the appearance of multiplicative and/or linear logarithmic corrections to the area law \cite{calabrese,vidallatorre03}.

Apart from the deep physical significance of this law, it has important implications regarding the possibility of simulating many-body quantum systems using tensor network (TN) states \cite{white92,schollwoeck04,murgadvances,ciracreview}.
For instance, it has been shown \cite{verstraetecirac05} that any state of a quantum spin system fulfilling the area law in one spatial dimension (including logarithmic violations) can be efficiently represented by a matrix product state (MPS) \cite{fannes92,perezgarcia06}, the simplest version of a TN.

Very recently, another remarkable discovery has been made with relation to the area law \cite{entspectrum}. It has been shown that for certain models in two
spatial dimensions, the reduced density matrix of a region $A$ has a very peculiar spectrum, which is called the "entanglement spectrum": by taking the logarithm of the eigenvalues of $\rho_A$, one obtains a spectrum that resembles very much the one of a 1-dimensional critical theory (i.e. as prescribed by conformal field theory).
This has been established for different systems as diverse as gapped
fractional quantum Hall states~\cite{entspectrum} or spin-1/2 quantum
magnets~\cite{poilblanc2010}. Interestingly, the correlation length in the bulk of the ground state can be naturally interpreted as a thermal length in
one dimension~\cite{poilblanc2010}.

This is all very suggestive for the fact that the reduced density matrix is the thermal state of a 1-dimensional theory. However, there is a clear mismatch in dimensions: the Hilbert space associated to $\rho_A$ has two spatial dimensions, while the 1-dimensional theory obviously has only 1. Intuitively, this is clear as all relevant degrees of freedom of $\rho_A$ should be located around the boundary of region $A$. The main question addressed in this paper is to explicitly identify the degrees of freedom on which this 1-dimensional Hamiltonian acts.

We show that projected entangled-pair states (PEPS) \cite{verstraetecirac04}
give a very natural answer to that question. The degrees of freedom of the 1-dimensional theory correspond to the virtual
particles which appear in the valence bond description of PEPS, and that
"live" at the boundary of region $A$
\cite{verstraetecirac04b,verstraetecirac04}. More specifically, PEPS are
built by considering a set of virtual particles at each node of the
lattice, which are then projected out to obtain the state of the physical
spins. As we show, the boundary Hamiltonian can be thought of as acting on
the virtual particles that live at the boundary of region $A$.
Furthermore, we will present evidence that, for gapped systems, such a boundary Hamiltonian is quasi--local (i.e. contains only
short-range interactions) in terms of those (localized) virtual particles. As a quantum phase transition is approached, the range of
the interactions increases. Finally, we will show that the interactions lose their local character for the case of quantum systems exhibiting topological order. We will also show how operators in the bulk can be mapped to operators on the boundary.

The fact that the boundary Hamiltonian is quasi--local has important
implications for the theory of PEPS which go well beyond those of the area
law. While PEPS are expected to accurately represent well the low energy
sector of local Hamiltonians in arbitrary dimensions \cite{hastings06}, it has not been proven that one can use them to
determine expectation values in an efficient and accurate way. For that,
one has to contract a set of tensors, a task which could in principle require
exponential time in the size of the lattice. In order to circumvent this
problem, a method was introduced \cite{verstraetecirac04} which successively approximates the boundary of a growing region by a matrix product density operator, which is exactly the density matrix of local virtual particles discussed before. It is not clear a
priori to which extent that density matrix can be approximated by a MPS; more specifically,
the bond dimension of that MPS could in principle grow exponentially with the size of the
system if a prescribed accuracy is to be reached, which would lead to an exponential scaling of the computational effort. However, that MPS does nothing but approximate the boundary density operator $\rho_A$ for different regions $A$. In case such an operator can
be written as a thermal state of a quasilocal Hamiltonian, it immediately
follows that in order to approximate it by a MPS one just needs a bond
dimension that scales polynomially with the lattice
size~\cite{hastings06}, and thus that expectation values of PEPS can be
efficiently determined.

\section{PEPS and boundary theories}

\subsection{Model}

\begin{figure}
\includegraphics[width=\columnwidth]{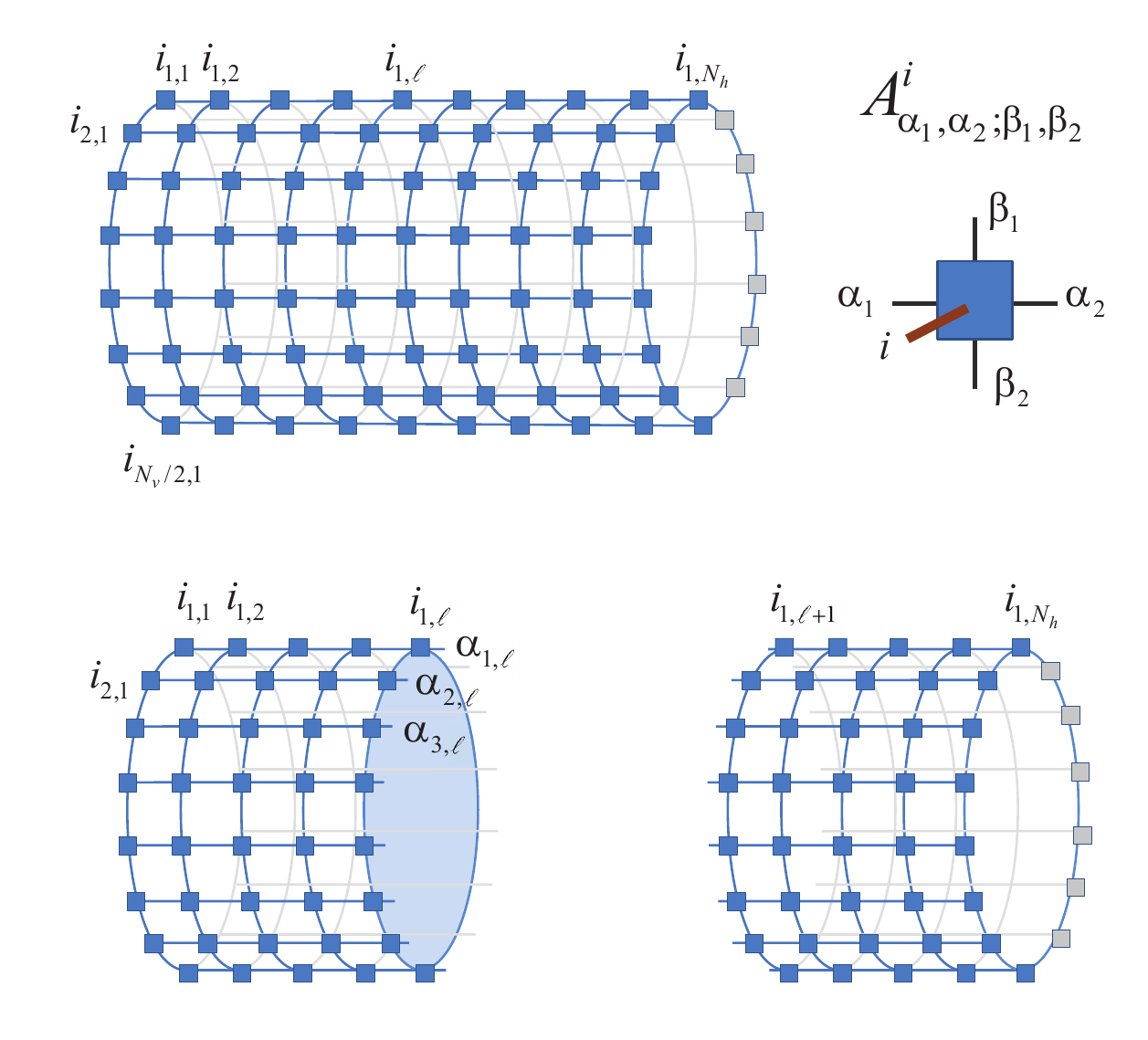}
\caption{Top: We consider an $N_v\times N_h$ spin lattice in a cylindrical geometry. The PEPS is obtained by replacing each lattice site with a tensor A, and contracting the virtual indices $\alpha$ and $\beta$ along the horizontal and vertical directions. Bottom: We cut the lattice into two pieces, left and right. The virtual indices $\alpha$ of the tensors $A$ along the cut are shown. The state $|\Psi_L\rangle$ acts on the spins ($i_{k,n}$) as well as on the virtual spins along the cut.}
\label{Fig1}
\end{figure}

We consider a PEPS, $|\Psi\rangle$, of an $N_v\times N_h$ spin lattice in two spatial dimensions. Note that one can always find a finite-range interaction Hamiltonian for which $|\Psi\rangle$ is a ground state \cite{verstraetewolf06}. We will assume that we have open (periodic) boundary conditions in the horizontal (vertical) direction: the spins are regularly placed on a cylinder and the state $|\Psi\rangle$ is translationally invariant along the vertical direction [see Fig. (\ref{Fig1})]. All spins have total spin $S$, except perhaps at the boundaries where we may choose a different spin in order to lift degeneracies related to the open boundary conditions. We will be interested in the reduced density operator, $\rho_\ell$, corresponding to the spins lying in the first $\ell$ columns; that is, when we trace all the spins from column $\ell+1$ to $N_h$.

More specifically, the effective Hamiltonian, $H_\ell$, corresponding to those spins,  is defined through $\rho_\ell=\exp(-H_\ell)/Z_\ell$, with $Z_\ell$ a normalization constant. We will be interested not only in the entanglement spectrum \cite{entspectrum}, but also in the specific form of $H_\ell$ and its interaction length, as we will define below.

In order to simplify the notation, it is convenient to label the spin indices of each column with a single vector. We define $I_n=(i_{1,n},i_{2,n},\ldots,i_{N_v,n})$, where $i_{k,n}=-S/2,-S/2+1,\ldots,S/2$ for $n=2,\ldots,N_h-1$ (for $n=1$ or $n=N_h$ we may have different spin $S$). Thus, we can write
 \be
 |\Psi\rangle = \sum_{I} c_{I}|I_1,I_2,\ldots,I_{N_h}\rangle.
 \ee
For a PEPS we can write
\be
 c_{I}= \sum_\Lambda L_{\Lambda_1}^{I_1} B_{\Lambda_1,\Lambda_2}^{I_2}
\ldots B_{\Lambda_{N_h-2},\Lambda_{N_h-1}}^{I_{N_{h-1}}}
R_{\Lambda_{N_h-1}}^{I_{N_h}} .
\ee
Here $\Lambda_n=(\alpha_{1,n},\alpha_{2,n},\ldots,\alpha_{N_v,n})$, where $\alpha_{k,n}=1,2,\ldots,D$ with $D$ the so-called bond dimension. Each of the $B^I$'s can be expressed in terms of a single tensor, $\hat A^i$,
 \be
 B_{\Lambda_{n-1},\Lambda_n}^{I_n} = \tr\left[\prod_{k=1}^{N_v} \hat A^{i_{k,n}}_{\alpha_{k,n-1},\alpha_{k,n}} \right],
 \ee
where for each value of $i,\alpha,\alpha'$, $\hat A^i_{\alpha,\alpha'}$ is a $D\times D$ matrix, with elements $A^i_{\alpha,\alpha';\beta,\beta'}$ (the indices $\alpha$ and $\beta$ correspond to the virtual particles entangled along the horizontal and vertical directions, respectively \cite{verstraetecirac04}; see Fig. \ref{Fig1}). For the first (left) and last (right) column we define $L^I$ and $R^I$ similarly in terms of the $D\times D$ matrices $\hat l^i_{\alpha}$, and $\hat r^i_{\alpha'}$:
 \begin{eqnarray}
 L_{\Lambda_1}^{I_1} &=& \tr\left[\prod_{k=1}^{N_v} \hat l^{i_{k,1}}_{\alpha_{k,1}} \right], \\
 R_{\Lambda_{N_h-1}}^{I_{N_h}} &=& \tr\left[\prod_{k=1}^{N_v} \hat r^{i_{k,N_h}}_{\alpha_{k,N_h-1}} \right].
 \end{eqnarray}
Thus, the tensors $\hat A$, $\hat l$, and $\hat r$ (for which explicit expressions will be given later on) completely characterize the state $|\Psi\rangle$, which is obtained by "tiling" them on the surface of the cylinder. The first has rank 5, whereas the other two have rank 4. Here we have taken all the tensors $A$ equal, but they can be chosen to be different if the appropriate symmetries are not present.

\subsection{Boundary density operator}

We now want to express the reduced density operator $\rho_\ell$ in terms of the original tensors. In order to do that, we block all the spins that are in the first $\ell$ columns, and those in the last $N_h-\ell$, and define
 \be
 \hat L^{I_a} = L^{I_1} B^{I_2} \ldots B^{I_\ell}, \quad \hat R^{I_b} = B^{I_{\ell+1}} \ldots B^{I_{N_h-1}} R^{I_{N_h}},
 \ee
where we have collected all the indices $I_1,\ldots,I_\ell$ in $I_a$ and the rest in $I_b$. With this notation, the state $|\Psi\rangle$ can be considered as a two-leg ladder, ie $\hat N_h=2$, and $\hat\ell=1$, where $\rho_{\hat\ell}$ is the density operator corresponding to a single leg. Thus, we have
 \be
 \label{2Leg}
 |\Psi\rangle = \sum_{I_a,I_b}\sum_{\Lambda} \hat L^{I_a}_\Lambda \hat R^{I_b}_\Lambda |I_a,I_b\rangle.
 \ee
It is convenient to consider the space where the vectors $L^I$ and $R^I$ act as a Hilbert space, and use the bra/ket notation there as well. That space, that we call {\it virtual space}, is the one corresponding to the ancillas that build the PEPS in the valence bond construction \cite{verstraetecirac04}. They are associated to the boundary between the $\ell$--th and the $\ell+1$st columns of the original spins. The dimension is thus $D^{N_v}$ (see Fig. \ref{Fig1}). In order to avoid confusion with the space of the spins, we have used $|v)$ to denote vectors on that space. We can define the (unnormalized) joint state for the first $\ell$ columns and the virtual space, $|\Psi_L\rangle$, and similarly for the last columns, $|\Psi_R\rangle$, as
 \be
 |\Psi_L\rangle = \sum_{I_a} |\hat L^{I_a})|I_a\rangle,\quad
 |\Psi_R\rangle = \sum_{I_a} |\hat R^{I_b})|I_b\rangle
 \ee
with
 \be
 |\hat L^{I_a})=\sum_\Lambda \hat L^{I_a}_\Lambda |\Lambda),\quad
 |\hat R^{I_b})=\sum_\Lambda \hat R^{I_b}_\Lambda |\Lambda),
 \ee
and $|\Lambda)$ the canonical orthonormal basis in the corresponding
virtual spaces.  The state $|\Psi\rangle$ can then be straightforwardly
defined in terms of those two states. The corresponding reduced density
operators for both virtual spaces are
 \be
 \label{sigmaLR}
 \sigma_L = \sum_{I_a} |\hat L^{I_a})(\hat L^{I_a}|\ ,\quad \sigma_R =
\sum_{I_b} |\hat R^{I_b})(\hat R^{I_b}|\ .
\ee
In terms of those operators, it is very simple to show that
\be
 \rho_\ell = \sum_{\Gamma,\Gamma'} \vert \chi_\Gamma\rangle\langle
\chi_{\Gamma'}| \; (\Gamma|\sqrt{\sigma_L^T} \sigma_R
\sqrt{\sigma_L^T}|\Gamma')
\ee
where $\vert\Gamma)$ is an orthonormal basis of the range of
$\sigma_L$, $\sigma_L^T$ is the transpose of $\sigma_L$ in the basis $|\Lambda)$,
and where we have defined an orthonormal set (in the spin space)
 \be
 |\chi_\Gamma\rangle = \sum_I (\Gamma|\frac{1}{\sqrt{\sigma_L}}|\hat L^I)
\;|I\rangle\ .
\ee
Now, defining an isometric operator that transforms the virtual onto the spin space
${\cal U}=\sum_\Gamma |\chi_\Gamma\rangle(\Gamma|$, we have
 \be
 \rho_\ell=U \sqrt{\sigma_L^T} \sigma_R \sqrt{\sigma_L^T} U^\dagger\ .
 \ee

The isometry $U$ can also be used to  map any operator acting on the bulk onto the virtual spin space; note that this map is an isometry and hence not injective, i.e. a boundary operators might correspond to many different bulk operators. This is of course a necessity, as $U$ is responsible for mapping a 2-dimensional theory to a 1-dimensional one.

\subsection{Boundary Hamiltonian}

The previous equation shows that $\rho_\ell$ is directly related to the
density operators corresponding to the virtual space of the ancillary
spins that build the PEPS. In particular, if we have
$\sigma_L^T=\sigma_R=:\sigma_b$ (eg., when we have the appropriate symmetries
as in the specific cases analyzed below), then $\rho_\ell=U \sigma_b^2 U^\dagger$. The reduced density operator $\rho_\ell$ is thus directly related to that of the virtual spins along the boundary. Since $U$ is isometric it conserves the spectrum and thus the entanglement spectrum of $\rho_\ell$ will coincide with that of $\sigma_b^2$.  By writing $\sigma_b^2=\exp(-H_b)$, we obtain an effective one-dimensional Hamiltonian for the virtual spins at the
boundary of the two regions  whose spectrum coincides with the
entanglement spectrum of $\rho_\ell$.

We will be interested to see to what extent $H_b$ is a local Hamiltonian
for the boundary (virtual) space. We can always write $H_b$ as a sum of
terms where different spin operators. For instance, for $D=2$, we can take the Pauli operators $\sigma_\alpha$ ($\alpha=x,y,z$) acting on different spins, and the identity operator on the rest. We group those terms into sums $h_n$, where each $h_n$ contains all terms with interaction range $n$, i.e.,
for which the longest contiguous block of identity
operators has length $N_v-n$.  For instance, $h_0$ contains only one term,
which is a constant; $h_1$ contains all terms where only one Pauli
operator appears; and $h_{N_v}$ contains all terms where no identity
operator appears. We define
 \be
 d_n=\tr(h_n^2)/2^{N_v}\, ,
 \label{Eq:dn}
 \ee
which expresses the strength of all the terms in the Hamiltonian with interaction length equal to $n$. A fast decrease of $d_n$ with $n$ indicates that the effective Hamiltonian describing the virtual boundary is quasi-local. In the examples we examine below this is the case as long as we do not have a quantum phase transition. In such a case, the length of the effective Hamiltonian interaction increases.

\subsection{Implications for PEPS}

In case $\sigma_b$ can be written in terms of a local boundary Hamiltonian one can draw important consequences for the theory of PEPS. In particular, it implies that the PEPS can be efficiently contracted, and correlation functions can be efficiently determined. The reason can be understood as follows. Let us consider again the cylindrical geometry (Fig.\ref{Fig1}), and let us assume that we want to determine any correlation function along the vertical direction, eg at the lattice points $(\ell,1)$ and $(\ell,x)$. It is very easy to show that such a quantity can be expressed in terms of $\sigma_L$ and $\sigma_R$. If we are able to write these two operators as Matrix Product Operators (MPO), ie as
 \be
 \sum_{i_n,j_n,=1}^D \tr\left[ M^{i_1,j_1} \ldots M^{i_{N_v},j_{N_v}}\right] |i_1,\ldots,i_{N_v}\rangle\langle j_1,\ldots,j_{N_v}|,
 \ee
where the $M'$ are $D'\times D'$ matrices, then the correlation function
can be determined with an effort that scales as $N_v (D')^6$. It has been
shown by Hastings \cite{hastings06} that if an operator can be written as
$\exp(-H_b/2)$, where $H_b$ is quasilocal, then it can be efficiently
represented by an MPO; that is, the bond dimension $D'$ only scales
polynomially with $N_v$. Thus, we have that the time required to
determine correlation functions only scales polynomially with $N_v$.

Later on, when we examine various examples, we will use MPO to represent
$\sigma_b$. In that case, we can directly check if we obtain a good
approximation by using a MPO just by simply observing how much errors
increase when we decrease the bond dimension $D'$. We will see that the
error increases when we approach a quantum phase transition. Furthermore,
whenever $\sigma_b$ can be well approximated by a MPO, we can use the
knowledge gained in the context of MPS \cite{fannes92,perezgarcia06} to
observe the appearance of a quantum phase transition in the original PEPS.
For that, we just have to recall that the correlation length, $\xi$, is
related to the two largest (in magnitude) eigenvalues, $\lambda_{1,2}$, of
the matrix $\sum_i M^{i,i}$; $\xi=1/\log(|\lambda_1/\lambda_2|)$. For
$|\lambda_1|=|\lambda_2|$, the correlation length diverges indicating the
presence of a quantum phase transition.

\subsection{Qualitative discussion}

In order to better understand the structure of $\sigma_b$, let us first consider a 1D spin chain. Even though the boundary of the chain, when cut into two parts, has zero dimensions, it will help us to understand the 2D systems. We take $N_v=1$ so that the PEPS reduces to a MPS. We can use the theory of MPS \cite{fannes92,perezgarcia06} to analyze the properties of the completely positive map (CPM) ${\cal E}$ (the matrices $A^i$ of the MPS are the Kraus operators of the CPM). In the limit $N_h\to\infty$, $\sigma_b$ is nothing but the fixed point of such a CPM. For gapped systems, ${\cal E}$ has a unique fixed point, and thus $\sigma_b$ is unique. For gapless systems, ${\cal E}$ becomes block diagonal (and thus there are several fixed points), the correlation length diverges, and we can write
 \be
 \label{gapless}
 \sigma_b=\oplus_{n=1}^B p_n \sigma_b^{(n)},
 \ee
where $B$ is the number blocks which coincides with the degeneracy of the eigenvalue of ${\cal E}$ corresponding to the maximum magnitude. In such case, the weights $p_n$ depend on the tensors $l$ and $r$ which are chosen at the boundaries. For critical systems, one typically finds that $D$ increases as a polynomial in $N_v$ such that one obtains logarithmic corrections to the area law \cite{Peschel,vidallatorre03}.

The 2D geometry considered here reduces to the 1D case if we take the
limit $N_h\to\infty$ by keeping $N_v$ finite. According to the discussion
above, we expect to have a unique $\sigma_b$ if we deal with a gapped system. As we will illustrate below with some specific examples, this operator can be
written in terms of a local Hamiltonian $H_b$ of the boundary virtual
space which is quasilocal. As we approach a phase transition, the gap
closes and the correlation length diverges. In some cases, the boundary density operator can  be written as a direct sum (\ref{gapless}), eventually leading to the loss of locality in the boundary Hamiltonian.

\section{Numerical Methods}

In order to determine $\sigma_b$ we make heavy use of the fact that $|\Psi\rangle$ is a PEPS.
We have followed three different complementary numerical approaches that we briefly describe here.

\subsection{Iterative procedure}

First of all, for sufficiently small values of $N_v$ (typically $N_v\le 12$) we can perform exact numerical calculations and determine $\sigma_{L,R}$ according to (\ref{sigmaLR}). The main idea is to start from the left and find first $\sigma_L$ for $\ell=1$ by contracting the tensors $l^i$ appropriately. Then, we can proceed for $\ell=2$ by contracting the tensors $A^i$ corresponding to the second column. In this vain, and as long as $N_v$ is sufficiently small we can determine $\sigma_{L,R}$ for all values of $\ell$ and $N_h$.

\subsection{Exact contractions and finite size scaling}

The second (exact) method is a variant applicable to larger values of $N_v$ (typically up to $N_v=20$) but restricted to a finite width in the
horizontal direction.
It consists in exactly contracting the internal indices of two adjacent blocks of size $N_v/2\times N_h$. These two blocks are then contracted together in a second step.
Although limited by the size $2^{N_v+2N_h}$ of the half-block (which has to fit in the computer RAM), this approach can still handle systems of size $20\times 2$ or $16\times 8$
and be supplemented by a finite size scaling analysis.

\subsection{Truncation method}

Finally, to take the $N_h\rightarrow \infty$ limit we can use the methods introduced in \cite{verstraetecirac04} to approximate the column operators. The main idea is to represent those operators by tensor networks with the structure of a MPS. We contract one column after each other, finding the optimal MPS after each contraction variationally. In particular, since we will consider translationally invariant states, we can choose the matrices of the corresponding MPS all equal, which simplifies the procedure. We can even approach the limit $N_v,N_h\to \infty$ as follows (see also \cite{murgadvances,vidal07}): (i) we start out with $\ell=1$, and contract the second column, obtaining another tensor network with the same MPS structure, but with increased bond dimensions. (ii) We continue adding columns, up to some $\ell=r$, where we start running out of resources. At that point, we have a tensor network with the MPS structure representing $\sigma_L$. Let us denote by $C^{n}_{\alpha,\beta}$ the basic tensor of that network, where $n=1,\ldots,D^2$ and $\alpha,\beta=1,\ldots,D^{2r}$ ($n$ denotes the index in the horizontal direction). (iii) When the bond indices $\alpha,\beta$ grow larger than some predetermined value, say $D_c\le D^{2r}$ we start approximating the tensor network by one with bond dimension $D_c$ as follows. We first construct the tensor $K_{\alpha,\alpha';\beta,\beta'}=\sum_n C^{n}_{\alpha,\beta}
\bar C^{n}_{\alpha',\beta'}$. Later on we will always deal with the case in which $K$ is hermitean (when considered as a matrix); if this is not the case, one can always choose a gauge where it is symmetric \cite{perezgarcia06}. We determine the eigenvector, $X_{\beta,\beta'}$, corresponding to the maximum eigenvalue of $K$, diagonalize $X$, consider the $D_c$ largest eigenvalues and build a projector onto the corresponding eigenspace. We then truncate the indices $\alpha$ and $\beta$ by projecting onto that subspace. (iv) We continue in the same vein until the truncated tensor structure converges, which corresponds to the limit $N_h\to\infty$. (v) We can do the same with $\sigma_R$ by going from right to left. For the examples studied below, $\sigma_L=\sigma_R=:\sigma_b=\sigma_b^T$, and thus we just have to carry out this procedure once.

\section{Numerical results for AKLT models}

\begin{figure}[h]\begin{center}
  \includegraphics[width=0.9\columnwidth]{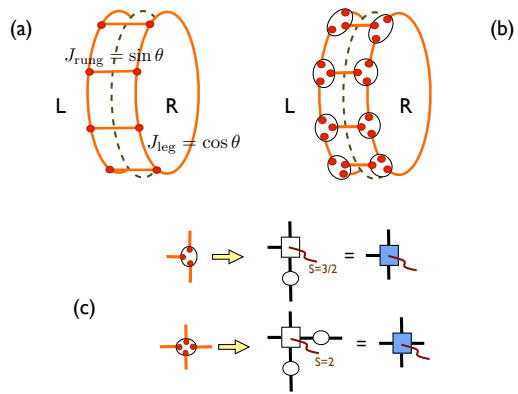}\end{center}
  \caption{(Color online)
(a) Ribbon made of two ($N_h=2$) coupled periodic S=1/2 Heisenberg chains (2-leg ladder).
(b) Groundstate of a 2-leg S=3/2 AKLT ladder.  Each site is split into
three spins-1/2 (red dots). Nearest neighbor spins-1/2 are paired up into
singlet valence bonds.  (c) PEPS representation for S=3/2 and S=2 sites
of AKLT wavefunctions in the valence bond (singlet) picture (for
connection to the "maximally entangled picture" see text). Open squares
stand for the $r^m_{\alpha_1,\alpha_2,\alpha_3}$ and
$A^m_{\alpha_1,\alpha_2,\alpha_3,\alpha_4}$ tensors defined in the text
and open circles correspond the to $2\times 2$ matrix $[0,1;-1,0]$.
}
\label{Fig:ladders}
\end{figure}

\begin{figure}[h]
\begin{center}
  \includegraphics[width=0.9\columnwidth]{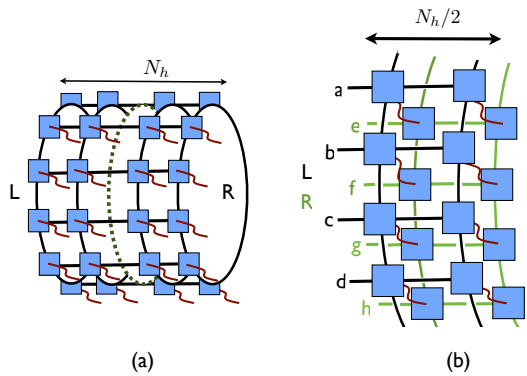}
\end{center}
\caption{(Color online)
(a) 4-leg ($N_h=4$) AKLT ladder on a cylinder partitioned (dotted green line) into two halves.
(b) Schematic representation of the density matrix $\sigma_b^2$ of a 4-leg ($N_h=4$, $\ell=2$) AKLT ladder. After being "cut" the two halves are "glued"  together (physical indices are contracted).
}
\label{Fig:4legs}
\end{figure}

We now investigate some particular cases. We concentrate on the AKLT model \cite{AKLT1,AKLT2}, whose ground state, $|\Psi\rangle$, can be exactly described by a PEPS with bond dimension $D=2$, as shown in Figs.~\ref{Fig:ladders} and  and \ref{Fig:4legs}. The spin in the first and last column have $S=3/2$, whereas the rest have $S=2$. The AKLT
Hamiltonian is given by a sum of projectors onto the subspace of maximum
total spin across each nearest neighbor pair of spins,
\be
 \label{AKLT}
 H_{\rm AKLT}=\sum_{<n,m>} P_{n,m}^{(s)}\ ,
 \ee
where $P_{n,m}^{(s)}$ is the projector onto the symmetric subspace of spins $n$ and $m$. This Hamiltonian is $su(2)$ and translationally invariant. This invariance is inherited by the virtual ancillas, and thus $\sigma_b$ and $H_b$ will also be. These symmetries can be used in the numerical procedures. Note that if $H_b$ has this symmetries and has short-range interactions, then since the ancillas have spin 1/2 (as $D=2$), it will be generically critical.

The lattice is bipartite. It is convenient to apply the operator
$\exp(i\pi S_y/2)$ to every spin on the B sublattice: this unitary
operator does not change the properties of $\rho_\ell$ but slightly
simplifies the description of the PEPS. Thus, we can write the AKLT
Hamiltonian as in (\ref{AKLT}) but with $P_{n,m}^{(s)}\to \tilde P_{n,m}
:= \exp(i\pi (f_n S_{y,n}+f_m S_{y,m}) P_{n,m}^{(s)}\exp(i\pi (f_n
S_{y,n}+f_mS_{y,m})$, with $f_n=0,1/2$ if the spin $n$ is in the A or B
sublattice, resp.

We will study finite $N_h$-legs ladders, as well as infinite square lattices.
We will start out in the next subsection with the simplest case of $N_h=2$. Note that for this particular case the subsystem we consider when we trace one of the legs is a spin chain itself, so that density operator $\rho_{\ell=1}$ already describes a 1-dimensional system and thus the physical spins already represent the boundary. In such a case, we do not need to resort to the PEPS formalism but we can also study other model Hamiltonians besides the AKLT one. For example, we will consider the $su(2)$-symmetric Heisenberg ladder Hamiltonian of $S=1/2$ [Fig. \ref{Fig:ladders}(a)]
\be
\label{Heis}
H_{\rm Heis}=\sum_{<n,m>} J_{n,m} {\bf S}_n\cdot {\bf S}_m\, ,
\ee
where the exchange couplings  $J_{n,m}$ are parametrized by some angle
$\theta$, i.e.\ $J_{\rm leg}=\cos{\theta}$ ($J_{\rm rung}=\sin{\theta}$) for
nearest-neighbor sites $n$ and $m$ on the legs (rungs) of the ladder.
Although the ground state has no simple PEPS representation, it can be
obtained numerically by standard Lanczos exact diagonalization techniques
on finite clusters of up to $14\times 2$ sites ~\cite{poilblanc2010}. Similarly to the AKLT 2--leg ladder [Fig. \ref{Fig:ladders}(b)], it possesses a finite magnetic correlation length $\xi$ which diverges when $\theta\rightarrow 0$ (decoupled chain limit). The opposite limit $\theta=\pi/2$ ($\theta=-\pi/2$) corresponds to decoupled singlet (triplet) rungs (strictly speaking, with zero correlation length).

For infinite systems, we will also be interested in the behavior of $H_b$ along a quantum phase transition. To this aim, we will also consider a distorted version of the AKLT model, and define a family of Hamiltonians
 \be
 H(\Delta)=\sum_{<n,m>} Q_n(\Delta)Q_m(\Delta) \tilde P_{n,m} Q_n(\Delta)Q_m(\Delta),
 \ee
where $Q_n(\Delta)=e^{-8\Delta S_{z,n}^2}$. Note that the Hamiltonian is translationally invariant and has $u(1)$ symmetry. As $\Delta$ increases, it penalizes (nematic) states with $S_z=0$, and thus the spins tend to
take their maximum value of $S_z^2$. As we will show, there exists a critical value of $\Delta$ where a quantum phase transition occurs.

\subsection{2--leg ladders~:~comparison between AKLT and Heisenberg models}

Let us start out with the $su(2)$-symmetric $\Delta=0$ AKLT model in a
two-leg ladder configuration, where $\rho_\ell$ corresponds to state of
one of the legs; that is, we take $N_h=2$, $\ell=1$, and all spins have
$S=3/2$ as shown in Fig.~\ref{Fig:ladders}(b). The
Hamiltonian is gapped~\cite{AKLT1,AKLT2}, and the ground state is a PEPS with
bond dimension $D=2$. The tensors corresponding to the two legs, $l$ and
$r$, coincide and are given by $r^m_{\alpha_1,\alpha_2,\alpha_3}= \langle
s_m|\alpha_1,\alpha_2,\alpha_3\rangle$, where $\alpha_i=\pm 1/2$, and
$|s_m\rangle$ is the state in the symmetric subspace of the three spin 1/2
with $S_z|s_m\rangle=m|s_m\rangle$, $m=-3/2,-1/2,1/2,3/2$.

\begin{figure}[h]
\begin{center}
  \includegraphics[width=0.98\columnwidth]{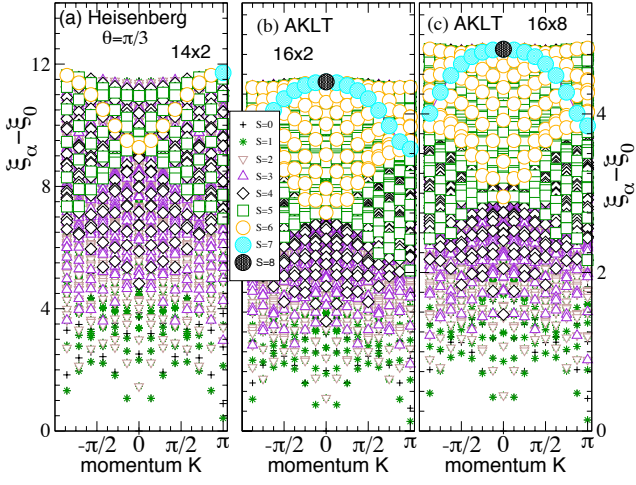}
\end{center}
\caption{(Color on line)
Entanglement spectra of $H_b$ (w.r.t. the groundstate energy $\xi_0$) versus total momenta $K$ in the chain (vertical) direction.
(a) 2--leg ($14\times 2$)
quantum Heisenberg ladder, (b) 2--leg ($16\times 2$) AKLT ladder and (c) 8--leg ($16\times 8$) AKLT ladder.
The eigenvalues are labelled according to their total spin quantum number using different symbols (according to
the legend on the graph).
}
\label{Fig:ES}
\end{figure}

We first examine the entanglement spectrum of $H_b$ computed on a
$16\times 2$ ladder.  It is shown on Fig.~\ref{Fig:ES}(b) as a function of
the momentum along the legs, making use of translation symmetry (the
vertical direction is periodic) enabling to block-diagonalize the reduced
density matrix in each momentum sector $K$. Note that it is also easy to implement the conservation of the $z$-component $S_z$ of the total spin so that each eigenstate can also be labelled according to its total spin $S$.  The
low-energy part of the spectrum clearly reveals zero-energy modes at $K=0$
and $K=\pi$ consistent with conformal field theory of central charge
$c=1$.

It is of interest to compare the 2--leg AKLT results to the ones of the 2-leg S=1/2 Heisenberg ladder (\ref{Heis}) sketched in Fig.~\ref{Fig:ladders}(a) and investigated in Ref.~\onlinecite{poilblanc2010}.
Fig.~\ref{Fig:ES}(a) obtained on a $14\times 2$ ladder for a typical
parameter $\theta=\pi/3$ shows the entanglement spectrum of $\rho_\ell$
which, again, is very similar to that of a single nearest-neighbor
Heisenberg chain. As mentioned in Ref.~\onlinecite{poilblanc2010}, in
first approximation, varying the parameter $\theta$ (and hence the ladder
spin-correlation length) only changes the overall scale of the energy
spectrum. Hence, it has been suggested~\cite{poilblanc2010} to connect
this characteristic energy scale to an effective inverse temperature
$\beta_{\rm eff}$.

\begin{figure}[h]
\begin{center}
  \includegraphics[width=0.9\columnwidth]{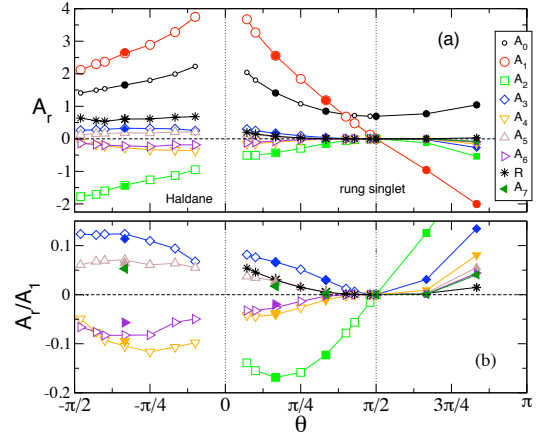}
\end{center}
\caption{(Color on line)
(a) Amplitudes $A_r$ of the (isotropic) spin-spin couplings up to distance $r=7$ of the effective boundary Hamiltonian
of a quantum Heisenberg 2-leg ladder in the Haldane and rung singlet phases vs $\theta$.
(b) Ratio of the same amplitudes normalized to the nearest-neighbor coupling ($r=1$).
Computations are carried out on $12\times 2$ (open symbols) and $14\times 2$ (closed symbols) systems.
Note that when $\theta\rightarrow \pi/2$ (decoupled rung singlets), $A_r/A_0 \rightarrow 0$ for $r\ge 1$ and all the
weights of the reduced density matrix becomes equal to $2^{-N_v}$ ($A_0=\ln{2}$).
}
\label{Fig:JvsTheta}
\end{figure}

The above results strongly suggest  that $H_b$ is "close" to a one-dimensional nearest-neighbor Heisenberg Hamiltonian. To refine this statement and make it more precise, we perform an expansion in terms of $su(2)$-symmetric
extended-range exchange interactions,
\be
H_b=A_0 N_v +\sum_{r,k} A_r  \, {\bf S}_k\cdot {\bf S}_{k+r} + R {\hat X} \, ,
\label{Eq:Hb}
\ee
where $R{\hat X}$ stands for the "rest", i.e.\ (small) multi-spin
interactions.  The amplitudes $A_r$ can be computed from simple trace
formulas,
\be
A_r= \frac{4}{N_v}{\rm tr}\{H_b\sum_k \sigma^z_k \sigma^z_{k+r}\} / 2^{N_v} \, ,
\ee
requiring the full  knowledge of the eigenvectors of $H_b$ (i.e.\ of
$\sigma_b$).  $A_0$ is fixed by some normalization condition, e.g. ${\rm
tr}\,\sigma_b=1$.  Assuming $\hat X$ is normalized as an extensive
operator in $N_v$, i.e.\ $\frac{1}{N_v}{\rm tr}\{{\hat X}^2 \}= 2^{N_v}$,
the amplitude $R$ is given by:
\be
R^2= \frac{1}{N_v}{\rm tr}\{H_b^2\}/2^{N_v}-N_v A_0^2-\frac{3}{16}\sum_{r=1}^{N_v/2} A_r^2 \, .
\ee

The coefficients $A_r$ and R of 2--leg Heisenberg ladders are plotted in
Fig.~\ref{Fig:JvsTheta}(a) as a function of the parameter $\theta$, both
in the Haldane ($J_{\rm rung}<0$ i.e. ferromagnetic) and rung singlet
phases ($J_{\rm rung}>0$ i.e. antiferromagnetic).
Generically, we find that $H_b$ is {\it not frustrated}, i.e.\ all
couplings at odd (even) distances are antiferromagnetic (ferromagnetic),
$A_r >0$ ($A_r <0$).  Clearly, the largest coupling is the
nearest-neighbor one ($r=1$).  Fig.~\ref{Fig:JvsTheta}(b) shows the
relative magnitudes of the couplings at distance $r>1$ w.r.t.\ $A_1$.
These data suggest that the effective boundary Hamiltonian $H_b$ is short
range, especially in the strong rung coupling limit
($\theta\rightarrow\pi/2$) where $|A_{r^\prime}/A_r|\rightarrow 0$ for
$r^\prime > r$. The amplitude $A_1$ of the nearest-neighbor interaction
can be identified to the effective inverse temperature $\beta_{\rm eff}$
which, therefore, vanishes (diverges) in the strong (vanishing) rung
coupling limit.

\begin{figure}[h]
\begin{center}
  \includegraphics[width=0.9\columnwidth]{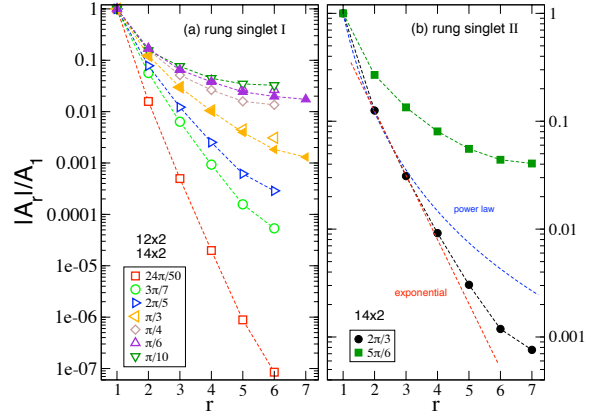}
\end{center}
\caption{(Color on line)
2-leg quantum Heisenberg ladders -- Ratio of the amplitudes $|A_r|$ by the
nearest-neighbor amplitude $A_1$ plotted using a logarithmic scale as a
function of $r$ for different values of $\theta$.  (a) Antiferromagnetic
and (b) ferromagnetic leg couplings (the rung couplings are
antiferromagnetic in both cases).}
\label{Fig:JvsDist}
\end{figure}

\begin{figure}[h]
\begin{center}
  \includegraphics[width=0.9\columnwidth]{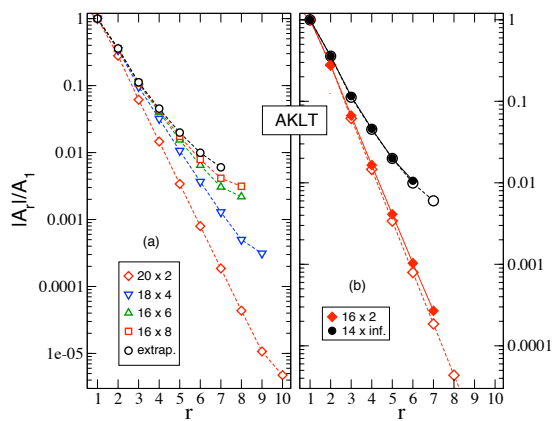}
\end{center}
\caption{(Color on line)
AKLT ladders -- (a) Ratio $|A_r|/A_1$ plotted using a logarithmic scale
as a function of $r$. Results are approximation-free for finite $N_h$ while the $N_h\rightarrow\infty$ limit
is obtained by finite size scaling (see Fig.~\protect\ref{Fig:JvsW}(b)). (b) Comparison with $\sqrt{d_{r+1}/d_2}$ (full symbols) computed (see text) on 2--leg and infinitely long ($N_h=\infty$)
cylinder.}
\label{Fig:JvsDist2}
\end{figure}

Next, we investigate the functional form of the decay of the amplitudes
$|A_r|$ with distance.  The ratio $|A_r|/A_1$ versus $r$ are plotted
(using semi-log scales) in Figs.~\ref{Fig:JvsDist}(a,b) for $12\times 2$
and $14\times 2$ Heisenberg ladders with different values of $\theta$.
Similar data for a $20\times 2$ AKLT ladder is shown in
Fig.~\ref{Fig:JvsDist2}(a), providing clear evidence of {\it exponential}
decay of the amplitudes with distance, i.e. $|A_r|\sim\exp{(-r/\xi_b)}$. The
Heisenberg ladder data are also consistent with such a behavior (even
though finite size corrections are stronger than for the AKLT case,
especially when $\theta\rightarrow 0$ or $\pi$).  It is not clear however
how deep the connection between the emerging length scale $\xi_b$ and the
2--leg ladder spin correlation length $\xi$ is. Note that the latter can
be related~\cite{poilblanc2010} to some effective {\it thermal} length
associated to the inverse temperature $\beta_{\rm eff}\propto A_1$.

Thanks to the PEPS representation of their ground state, AKLT ladders can be
(exactly) handled up to larger sizes than their Heisenberg counterparts
(typically up to $N_v=20$) enabling a careful finite size scaling analysis
of the boundary Hamiltonian (\ref{Eq:Hb}).  As shown in
Fig.~\ref{Fig:JvsW}(a), we observe a very fast (exponential) convergence
of the coefficients $A_r$ with the ladder length $N_v$. Hence, one gets at
least 7 (3) digits of accuracy for all distances up to $r=5$ ($r=7$).

In fact, as pointed out previously, the boundary Hamiltonian $H_b$ should not contain only two-body spin interactions.
However, the total magnitude of all left-over (multi-body) contributions, $R$, is remarkably small in the
AKLT 2--leg ladder~: as shown in Fig.~\ref{Fig:JvsW}(a), $R<A_4$.
In fact, the full magnitude of {\it all} many-body terms extending on $r+1$ sites is
given by $\sqrt{d_{r+1}}$ and can be compared directly to $|A_r|$ (after proper normalization).
Fig.~\ref{Fig:JvsDist2}(b) shows that $\sqrt{d_{r+1}/d_2}$ and $|A_r|/A_1$ are quite close, even at large distance.
Note however that multi-body interactions are significantly larger in the boundary Hamiltonian of
the Heisenberg ladder, as shown in Fig~\ref{Fig:JvsTheta} (although no
accurate finite size scaling analysis can be done in that case).

\subsection{$N_h$--leg AKLT ladder}

Now we consider the AKLT model on an $N_h$--leg ladder configuration; we
take $\ell=N_h/2$. The spins in the first and last legs have $S=3/2$, and the corresponding tensors coincide with the ones given above. The rest of the
spins have $S = 2$, and the corresponding tensor is $A^m_{\alpha_1,\alpha_2,\alpha_3,\alpha_4}= \langle
 s_m|\alpha_1,\alpha_2,\alpha_3,\alpha_4\rangle$,
where $\alpha_i=\pm 1/2$, and $|s_m\rangle$ is the state in the symmetric
subspace of the four spin 1/2 with $S_z|s_m\rangle=m|s_m\rangle$,
$m=-2,-1,0,1,2$ (see Fig.~\ref{Fig:ladders}(c)). An example of a 4--leg ladder and of a schematic representation of $\rho_\ell$ is shown in
Fig.~\ref{Fig:4legs}.

\begin{figure}[h]
\begin{center}
  \includegraphics[width=0.9\columnwidth]{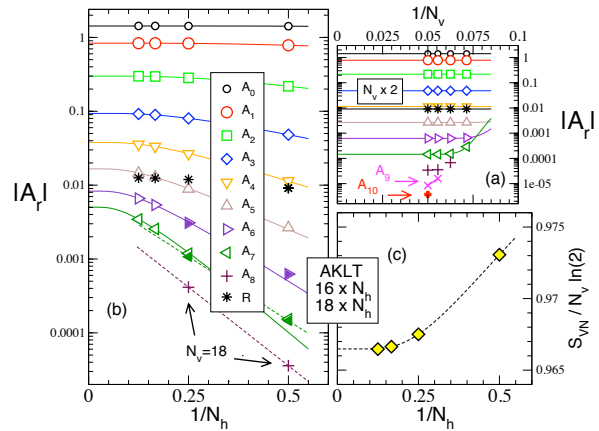}
\end{center}
\caption{(Color on line)
(a) Finite size scaling of the amplitudes $A_r$ for a 2-leg AKLT ladder vs $1/N_v$ ($N_v$=14, 16, 18, 20).
(b) Finite size scaling of the amplitudes $A_r$ for $N_h$-leg AKLT ladders vs $1/N_h$ at fixed
$N_v=16$ (open symbols) or $N_v=18$  (filled and $+$ symbols).
(c) VN entropy per unit length (normalized by $\ln(2)$) vs $1/N_h$ at fixed $N_v=16$.
}
\label{Fig:JvsW}
\end{figure}

Let us now follow the same analysis (\ref{Eq:Hb}) of the boundary
Hamiltonian as we did for the case of 2 legs.  The decay with distance of
the coefficients $A_r$ are reported in Fig.~\ref{Fig:JvsDist2}(a) for
4--leg, 6--leg and 8--leg AKLT ladders. Clearly, the decay is still
exponential with distance for all values of $N_h$ studied but the
characteristic length scale associated to this decay (directly given by
the inverse of the slope of the curve in such a semi-log plot) smoothly
increases with $N_h$.  A careful finite size scaling is performed in
Fig.~\ref{Fig:JvsW}(b)  to extract the $N_h\rightarrow\infty$ limit of all
$A_r$ (accurate up to $r=7$). The extrapolated values are reported in
Fig.~\ref{Fig:JvsDist2}(a) showing that $A_r$ also decays exponentially
fast with $r$ in an infinitely long cylinder ($N_h=\infty$).  The
characteristic emerging length scale is estimated to be still very short
around $1$.

Lastly, we compute the Von Neumann entanglement entropy defined by $S_{\rm VN}(\rho_\ell)=-{\rm tr} \{\rho_\ell \ln{\rho_\ell}\}$
with the normalization ${\rm tr}\,\rho_\ell=1$.
$S_{\rm VN}$ scales like $N_v$ ("area" law) and is bounded by $N_v \ln{2}$. Fig.~\ref{Fig:JvsW}(c) shows that the entropy
converges very quickly with $N_h$ to its thermodynamic value which is very close to the maximum value.
The entanglement of the two halves of the AKLT cylinder is therefore very strong.

\subsection{Thermodynamic limit and phase transitions}

Now we consider the $N_v,N_h\to\infty$ for the deformed AKLT model in order to investigate the phase transition. We will compare some of the results with the $2$--leg ladder as well. The spins in the first and last legs have $S=3/2$, and the rest $S=2$. The corresponding tensors are defined according to
 \begin{eqnarray}
 l^m_{\alpha_1,\alpha_2,\alpha_3} &=& r^m_{\alpha_1,\alpha_2,\alpha_3}= \langle s_m|Q(\Delta)|\alpha_1,\alpha_2,\alpha_3\rangle,\nonumber\\
 A^m_{\alpha_1,\alpha_2,\alpha_3,\alpha_4}&=& \langle
 s_m|Q(\Delta)|\alpha_1,\alpha_2,\alpha_3,\alpha_4\rangle,
 \end{eqnarray}
where $\alpha_i=\pm 1/2$, and $|s_m\rangle$ is the state in the symmetric
subspace of the three (four) spin 1/2 with $S_z|s_m\rangle=m|s_m\rangle$,
$m=-3/2,-1/2,1/2,3/2$ ($m=-2,-1,0,1,2$), respectively. 

\begin{figure}[h]
\begin{center}
  \includegraphics[width=0.9\columnwidth]{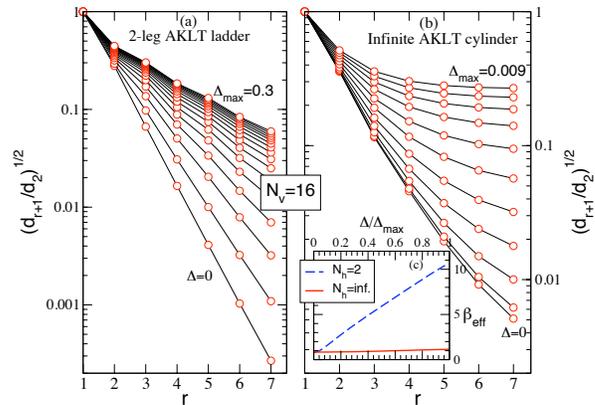}
\end{center}
\caption{(Color on line)
AKLT model with finite "nematic" field $\Delta$ -- (a) Relative amplitude $\sqrt{d_{r+1}/d_2}$ in a 2--leg ladder
plotted using a logarithmic scale as a function of $r$; (b) same for an infinitely long
cylinder ($N_h=\infty$). From bottom to top, $\Delta$ is incremented from $0$ to $\Delta_{\rm max}$
by constant steps.
(c) Inset~: effective temperature $\beta_{\rm eff}$ (see Eq.~\protect\ref{Eq:beta}) versus $\Delta$ for the two cases reported in (a) and (b).
All results are obtained for $N_v=16$ ($D_c=50$, and 100 iterations such that the tensors $C$ already converge). }
\label{Fig:dn}
\end{figure}

We will use the approximate procedure sketched in Section III-C. In particular, for $N_v$ larger than the correlation length the obtained tensors $C^{n}_{\alpha,\beta}$ will be independent of $N_v$. We have considered those tensors (with $D_c=50$ and 100 iterations), and built $\sigma_b$ and $H_b$ out of them.  Note that the $su(2)$ symmetry is explicitly broken by a finite $\Delta$ so that it becomes more convenient to use the variable $d_n$ of Eq.~(\ref{Eq:dn}) instead of $A_r$ to probe the spatial extent of $H_b$.  We recall that $(d_n)^{1/2}$ is the
mean amplitude of {\it all} interactions acting at distance $r=n-1$.  We
have plotted in Fig.~\ref{Fig:dn} all $d_n$, $n\le N_v/2$, for $N_v=16$ as
a function of $\Delta$.  As $\Delta$ increases, we see that the
interaction length of the effective Hamiltonian increases and one sees a
long-range interaction appearing. This indicates that we approach a phase
transition.  For the case of the ladder, the interaction length remain
practically constant for the same range of variation of $\Delta$.

Similarly to the investigation of the Heisenberg ladder~\cite{poilblanc2010}, it is interesting to define an
effective inverse temperature via the amplitude of the nearest-neighbor interaction,
\be
\beta_{\rm eff}=8\sqrt{\frac{d_2}{3}}\, ,
\label{Eq:beta}
\ee
where the pre-factor is introduced conveniently so that $\beta_{\rm eff}=A_1$ in the $su(2)$-symmetric limit $\Delta=0$.
As seen in the inset of Fig.~\ref{Fig:dn}, the inverse temperature of the ladder scales linearly with $\Delta$.
For the infinite cylinder, no singularity of $\beta_{\rm eff}$ is seen at the cross-over between short and long-range
interactions.

\begin{figure}[h]
\begin{center}
  \includegraphics[width=0.9\columnwidth]{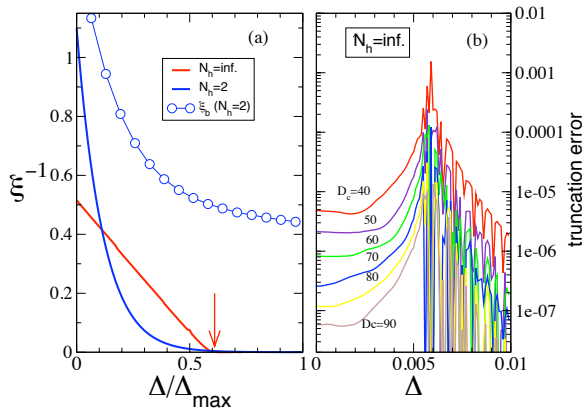}
\end{center}
\caption{(Color online)
(a) Inverse correlation length $\xi^{-1}$ vs $\Delta$ for both 2-leg ladder and infinitely long cylinder ($N_h=\infty$).
These data correspond to the infinite circumference limit, i.e. $N_v=\infty$. The arrow marks the phase transition in the infinitely
long cylinder.
Comparison with the inverse of the "emerging" length scale $\xi_b$ obtained by fitting the decay of the coefficients of $H_b$
plotted in Fig.~\ref{Fig:dn}(a) as $\sqrt{d_{r+1}/d_2}\sim \exp{(-r/\xi_b)}$.
(b) Truncation error in the $N_h\rightarrow\infty$ procedure. The results are compared with those obtained with $D_c=150$, and 100 iterations have always been used.
 }
\label{Fig:xi}
\end{figure}

Next, we plot the inverse correlation length as a function of $\Delta$
both for one dimension (i.e.\ an infinitely long ladder) and for two
dimensions (i.e.\  $N_v=N_h=\infty$) in Fig.~\ref{Fig:xi}(a), obtained with $D_c=150$ and 100 iterations (no difference are observed by taking $D_c=50$ and 50 iterations).  Clearly, the divergence of $\xi$ (i.e.\ $\xi^{-1}=0$) shows the appearance of a phase transition at $\Delta=0.0061$ in two dimensions. In contrast, $\xi^{-1}$ never crosses zero in the case of the ladder (i.e.\ in one dimension).  We have compared $\xi$ with the "emerging" length scale $\xi_b$ obtained by fitting the decay of the coefficients of $H_b$ as $\sqrt{d_{r+1}/d_2}\sim \exp{(-r/\xi_b)}$ on $N_v=16$ 2-leg and infinitely long (i.e.\ $N_h=\infty$) cylinders.  In the two leg ladder, we see that the
divergence of the correlation length $\xi$ for $\Delta\rightarrow\infty$
results from the interplay between (i) a (moderate) increase of the range
$\xi_b$ of the Hamiltonian $H_b$ and (ii) a linear increase with $\Delta$
of the effective temperature scale $\beta_{\rm eff}$, therefore approaching
the $T_{\rm eff}\rightarrow 0$ limit when $\Delta\rightarrow\infty$.  This
contrasts with the case of two dimensions ($N_v=N_h=\infty$) where the
divergence of $\xi$  occurs at {\it finite effective temperature} when
$H_b$ becomes "sufficiently" long-range.  It is however hazardous to fit
the decay of the coefficients of $H_b$ to obtain its functional form at
the phase transition.  Finally, in Fig.~\ref{Fig:xi}(b) we have plotted
the truncation error made by taking different $D_c$ in the limit
$N_h\rightarrow\infty$, and, again around $\Delta\approx 0.006$ the error increases. This is consistent with the expectation that as $H_b$ contains longer range interaction, the boundary density operator $\sigma_b$ requires a higher bond dimension to be described as a TN state.

\section{Numerical results for Ising PEPS}

We now continue by considering the Ising PEPS introduced in \cite{verstraetewolf06}. They all have bond dimension $D=2$ and
exhibite the $\mathbb Z_2$-symmetry of the transverse Ising chain. They depend on a single parameter, $\theta\in [0,\pi/4]$. For $\theta\sim \pi/4$ one has a state with all the spins pointing in the $x$ direction, whereas for
$\theta\sim 0$ the state is of the GHZ type (a superposition of all spins up and all down). In the thermodynamic limit ($N_v,N_h\to\infty$) for $\theta\approx 0.35$ they feature a phase transition, displaying critical behavior, where the correlation functions decay as a power law. Thus, by changing $\theta$ we can investigate how the boundary Hamiltonian behaves as one approaches the critical point.

\subsection{2--leg ladders}

The tensors corresponding to the two legs, $l$ and $r$, coincide and are given by $r^m_{\alpha_1,\alpha_2,\alpha_3}=
a_m(\alpha_1)a_m(\alpha_2)a_m(\alpha_3)$, where $m=0,1$, $\alpha_i=\pm
1/2$ and $a_m(\alpha)$ are parametrized as
$a_0(-1/2)=a_1(1/2)=\cos{\theta}$ and $a_0(1/2)=a_1(-1/2)=\sin{\theta}$.

\begin{figure}[h]
\begin{center}
  \includegraphics[width=0.9\columnwidth]{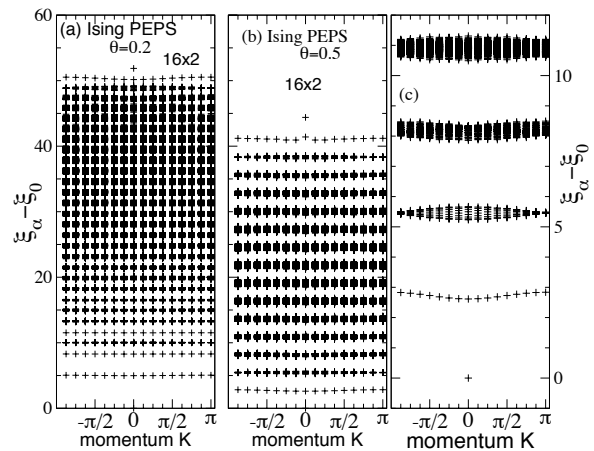}
\end{center}
\caption{
Entanglement spectrum of a $16\times 2$ Ising PEPS ladder versus momentum along
the ladder leg direction. Comparison between $\theta=0.2$ (a) and
$\theta=0.5$ (b) using the same energy scale. (c) Zoom in of the low-energy part of (b). }
\label{Fig:ES_Ising}
\end{figure}

As seen in Fig.~\ref{Fig:ES_Ising} the entanglement spectrum of the 2-leg
ladder is gapped for all $\theta$ values and resemble the one of an Ising
chain (equally spaced levels) with small quantum fluctuations revealed by
the small dispersion of the bands.  The effective inverse temperature,
qualitatively given by the gap (or the spacing between the bands),
decreases for increasing $\theta$.

The interaction length of the boundary Hamiltonian for the ladder is displayed in Fig.~\ref{Fig:dnIsing}(a). The strength of the interactions decay exponentially with the distance for all values of $\theta$. As we increase this angle, one only observes a decrease of the interaction length. Note that as opposed to the AKLT models studied in the previous sections, $d_1\ne 0$.  Indeed, there always exists a term with a single Pauli operator $\sigma_x$, describing an effective transverse field in $H_b$. Thus, that Hamiltonian is given by a transverse Ising chain in the non--critical region of parameters.

We have also plotted the inverse correlation length $\xi^{-1}$ as a function of $\theta$ in Fig.~\ref{Fig:xi_Ising} (blue empty dots). While the correlation length increases as $\theta$ decreases, it only tends to infinite in the limit $\theta\to 0$, as it must be for a GHZ state. No signature of a phase transition is found otherwise.

\subsection{Thermodynamic limit and phase transitions}

We now move to the case of an infinitely long cylinder. As above to grow the cylinder in the horizontal direction, one
considers rank-5 tensors, which here take the form $A^m_{\alpha_1,\alpha_2,\alpha_3,\alpha_4}= a_m(\alpha_1)a_m(\alpha_2)
a_m(\alpha_3)a_m(\alpha_4)$, and use the same approximation scheme with 100 iterations as before.

\begin{figure}[h]
\begin{center}
  \includegraphics[width=0.9\columnwidth]{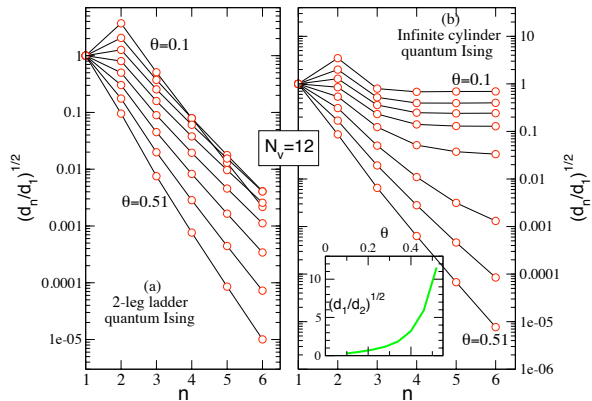}
\end{center}
\caption{(Color online)
Ising PEPS -- Relative amplitude $\sqrt{d_{n}/d_1}$ in a 2--leg ladder (a)
and in an infinitely long ($N_h=\infty$) cylinder (b) as a function of $n$ for $\theta$ varying from 0.1 to $\sim 0.51$ with constant intervals
(0.1,    0.1585,    0.2171,    0.2756,    0.3342,    0.3927,    0.4512 and 0.5098, from top to bottom).
Logarithmic scales are used on the vertical axis in both (a) and (b).
Inset: Ratio of the effective transverse field $\sqrt{d_1}$ over the effective Ising nearest-neighbor coupling $\sqrt{d_2}$ versus
$\theta$ for $N_h=\infty$ ($D_c=50$ and 100 iterations).  All results are obtained for $N_v=12$. }
\label{Fig:dnIsing}
\end{figure}

The parameters $d_n$ describing the boundary Hamiltonian $H_b$ behave very differently in the ladder and infinite cylinders as shown in Fig.~\ref{Fig:dnIsing}. While for the Ising PEPS ladder
$H_b$ remains short-ranged with exponential decay of $d_n$ vs $n$, the infinite cylinder shows a transition towards long-range interactions suggesting the existence of a phase transition. This is very similar to what occurred in the AKLT distorted model.

\begin{figure}[h]
\begin{center}
  \includegraphics[width=0.9\columnwidth]{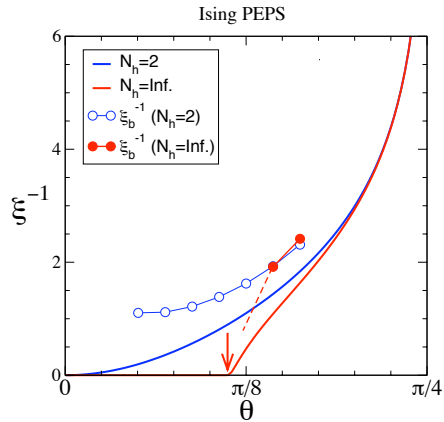}
\end{center}
\caption{(Color on line)
Ising PEPS --
Inverse correlation length $\xi^{-1}$ vs $\Delta$ for both 2-leg ladder and infinitely long cylinder ($N_h=\infty$, $D_c=150$ and 100 iterations).
These data correspond to the infinite circumference limit, i.e. $N_v=\infty$. The arrow marks the phase transition in the infinitely
long cylinder.
Comparison with the inverse of the "emerging" length scale $\xi_b^{-1}$ obtained by fitting the decay of
the coefficients plotted in Fig.~\protect\ref{Fig:dnIsing} as $\sqrt{d_{n}}\sim \exp{(-n/\xi_b)}$. }
\label{Fig:xi_Ising}
\end{figure}

As long as $H_b$ remains short-range, the density matrix $\rho_\ell$ can be
(qualitatively) mapped onto the thermal density matrix of an effective
quantum Ising chain (including a "family" of transverse-like fields) and,
therefore, no ordering is expected (at finite effective temperature).  A
phase transition however can appear when $H_b$ becomes long-ranged as it
is the case for an infinitely long cylinder.  This is evidenced by the the
behavior of correlation lengths computed for the 2-leg and infinitely long
($N_h=\infty$) cylinder and reported in Fig.~\ref{Fig:xi_Ising}. These
correlation lengths are compared to the respective "emerging" length
scales $\xi_b$ characterizing the decay of $\sqrt{d_{n}}$ with $n$.  In
the 2-leg ladder case, $\xi_b$ increases quite moderately when
$\theta\rightarrow 0$ ($\xi_b\sim 1$) so that  the divergence of the
correlation length $\xi$ in this limit is only attributed to a {\it
vanishing} of the effective temperature scale $T_{\rm eff}$. In contrast,
as for the AKLT model, the phase transition in two dimensions occurs at
finite (effective) temperature at the point where
$\xi_b\rightarrow\infty$.

In summary, these results evidence that whenever we approach a phase transition, the interaction length of the boundary Hamiltonian increases.

\section{Topological Kitaev code}

\begin{figure}[h]
\begin{center}
  \includegraphics[width=0.9\columnwidth]{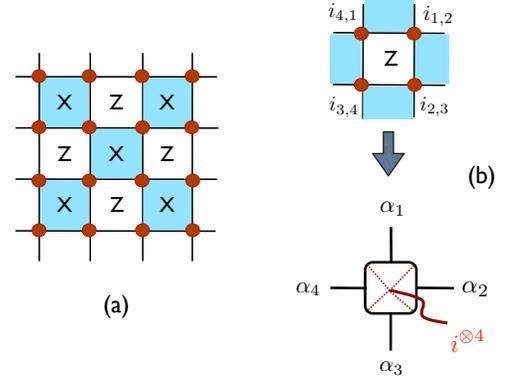}
\end{center}
\caption{(Color on line)
(a) Checkerboard decomposition in the Kitaev code. Spin-$\tfrac12$ are represented by (red) dots
at the vertices of the square lattice. $X$ and $Z$ operators act on the 4 spins of each type of (shaded and non-shaded) plaquettes.
(b) PEPS  representation of the Kitaev code (see text).}
\label{Fig:kitaev}
\end{figure}

Let us finally consider systems with topological order. We will focus on
Kitaev's code state~\cite{kitaev:toriccode}: It can be defined on a square
lattice with spin-$\tfrac12$ systems (qubits) on the vertices, with two
types of terms in the Hamiltonian,
\begin{equation}
\label{eq:tcode-terms}
h_X = X^{\otimes 4}\ , \quad
h_Z= Z^{\otimes 4}
\end{equation}
(where $X$ and $Z$ are Pauli matrices), each of which acts on the four
spins adjacent to a plaquette, and where the $h_X$ and $h_Z$ form a
checkerboard pattern (see Fig.~\ref{Fig:kitaev}(a)).  The ground state subspace of the code state can be
represented by a PEPS with $D=2$~\cite{verstraetewolf06}; a particularly
convenient representation is obtained by taking $2\times 2$ blocks of
spins across $h_Z$ type plaquettes, and jointly describing the spins in
each block by one tensor of the form~\cite{schuch:peps-sym}
\begin{equation}
\label{eq:tcode-tensor}
A^{i_{1,2},i_{2,3},i_{3,4},i_{4,1}}_{\alpha_1,\alpha_2,\alpha_3,\alpha_4}
 = \left\{\begin{array}{ll}
    1 &\mathrm{if\ } i_{x,x+1} = \alpha_{x+1} - \alpha_{x}
	\ \mathrm{mod}\,2 \ \forall\, x\\
    0 & \mathrm{otherwise.}
    \end{array}\right.
\end{equation}
Here, $i_{x,x+1}$ denotes the spin located between the bonds $\alpha_x$
and $\alpha_{x+1}$ (numbered clockwise) as shown in Fig.~\ref{Fig:kitaev}(b). It can be checked
straightforwardly that the resulting tensor network is an eigenstate of
the Hamiltonians of Eq.~(\ref{eq:tcode-terms}). Excitations of the model
correspond to violations of $h_X$-terms (charges) or $h_Z$-terms (fluxes),
which always come in
pairs~\cite{kitaev:toriccode}.

We put the code state on a cylinder of $N_h\times N_v$ tensors (i.e.,
$2N_h\times 2N_v$ sites), where we choose boundary conditions
\begin{equation}
\label{eq:kitaev-bnd}
|\chi_\theta) = \cos\tfrac\theta2\,|0)^{\otimes N_v}+
    \sin\tfrac\theta2\,|1)^{\otimes N_v}\ .
\end{equation}
This yields a state which is also a ground state of
$h_Z^b=Z^{\otimes 2}$ terms at the boundary, but not of the corresponding
$X^{\otimes 2}$ boundary terms; in other words, charges (Pauli $Z$ errors)
can condense at the boundaries of the cylinder~\cite{dennis}.
The full Hamiltonian---including the $h_Z^b$ terms at the boundary---has a
two-fold degenerate ground state which is topologically protected, and
where the logical $X$ and $Z$ operators are a loop of Pauli $X$'s around
the cylinder and a string of Pauli $Z$'s between its two ends (where they
condense), respectively.

To compute $\rho_\ell$, we start by considering the PEPS on the cylinder
without the boundary conditions (\ref{eq:kitaev-bnd}), i.e., with open
virtual indices at both ends (labelled $B$ and
$B'$). Cutting the cylinder in the middle leaves us with with
$\sigma_{BL}$, the joint reduced density operator for the virtual spaces
at the boundary, $B$ (or $B'$),  and the cut, $L$ (or $R$). From
(\ref{eq:tcode-tensor}), one can readily infer that the transfer operator
for a single tensor is $\openone^{\otimes 4}+X^{\otimes 4}$, and thus,
\begin{equation}
        \label{eq:tcode-1}
\sigma_{BL}=\sigma_{B'R} \propto
    \openone^{\otimes N_v}\otimes\openone^{\otimes N_v} +
        X^{\otimes N_v}\otimes X^{\otimes N_v}\ ,
\end{equation}
where the two tensor factors correspond to the $B$ ($B'$) and $L$ ($R$)
boundary, respectively.  Imposing the boundary condition
$|\chi_\theta)(\chi_\theta|$, Eq.~(\ref{eq:kitaev-bnd}), at $B$ ($B'$),
we find that (up to normalization)
\[
\rho_\ell \propto
    (1+\sin^2\theta)\,\openone^{\otimes N_v} +
    (2\sin\theta)\, X^{\otimes N_v}\ ,
\]
which is the thermal state $\rho_\ell\propto \exp[-\beta_\mathrm{eff}
H_\ell]$ of $H_\ell=-\mathrm{sign}(\sin\theta)\,X^{\otimes N_v}$ at
an effective inverse temperature
\[
\beta_\mathrm{eff} = \left\vert
    \tanh^{-1}\left[\frac{2\sin\theta}{1+\sin^2\theta}\right]
    \right\vert\ .
\]
The fact that $H_\ell$ acts globally is a signature of the topological
order, and comes from the fact that measuring an $X$ loop operator gives a
non-trivial outcome (namely $\sin\theta$).  Note that the entropy
$S(\rho_\ell)$ increases by one as $1/\beta_\mathrm{eff}$ goes from zero
to infinity.  This can be understood as creating an entangled pair of
charges $\ket{\mathrm{vac}}+f(\beta_\mathrm{eff})\ket{c,c^*}$ across the
cut, thereby additionally entangling the two sides by at most an ebit, and
subsequently condensing the charges at the boundaries.

Instead of considering $\sigma_L$, one can also see the topological order
by looking at $\sigma_{BL}$: It is the zero-temperature state of a
completely non-local Hamiltonian $X^{\otimes N_v}\otimes X^{\otimes N_v}$
which acts simultaneously on both boundaries in a maximally non-local way;
this relates to the fact that the expectation values of any two $X$ loop
operators around the cylinder are correlated.

Let us point out that systems with conventional long-range order behave
quite differently, even though they also exhibit correlations between
distant boundaries. Consider the spin-$\tfrac12$ Ising model
without field, which has a PEPS tensor
\[
A^i_{\alpha_1,\alpha_2,\alpha_3,\alpha_4}=
    \delta_{i,\alpha_1}\delta_{\alpha_1,\alpha_2}
    \delta_{\alpha_2,\alpha_3}\delta_{\alpha_3,\alpha_4}\ .
\]
The resulting local transfer operator is $|0)(0|^{\otimes 4}+|1)(1|^{\otimes
4}$, and thus,
\[
\sigma_{BL}=
|0)(0|^{\otimes N_v} + |1)(1|^{\otimes N_v}\ .
\]
By imposing boundary conditions at $B$, one
arrives at
\[
\rho_\ell =
\sin\theta\,|0)(0|^{\otimes N_v}
+ \cos\theta\,|1)(1|^{\otimes N_v}\ ,
\]
which is the thermal state of of the classical Ising Hamiltonian
\[
H(\beta) = -\sum_{i} Z_i Z_{i+1} -
\frac{\log\tan\theta}{2\beta N_v} \sum_i Z_i
\]
for $\beta\rightarrow\infty$. Thus, for the Ising model, $\rho_\ell$
is described by a local Ising Hamiltonian, rather than a completely
non-local interaction as for Kitaev's code state. The same holds true for
$\sigma_{BL}$, which is the ground state of a classical Ising model
without field: while it has correlations between the two boundaries, they
arise from a local (i.e., few-body) interaction coupling the two
boundaries, rather than from terms acting on \emph{all} sites on both
boundaries together.  Correspondingly, the long-range correlations in the
Ising model can be already detected by measuring local observables, instead of
topologically nontrivial loop operators as for Kitaev's code state.

\section{Conclusions and outlook}

In this paper, we have introduced a framework which allows to associate
the bulk of a system with its boundary in the spirit of the holographic
principle. To this end, we have employed the framework of Projected
Entangled Pair States (PEPS) which provide a natural mapping between the
bulk and the boundary, where the latter is given by the virtual degrees of
freedom of the PEPS.  This framework allows to map the state of any region
to a Hamiltonian on its boundary, in such a way that the properties of the
bulk system, such as entanglement spectrum or correlation length, are
reflected in the properties of the
Hamiltonian.  Since our framework also identifies observables in the bulk
with observables on the boundary, it establishes a general holographic
principle for quantum lattice systems based on PEPS.

In order to elucidate the connection between the bulk system and the
boundary Hamiltonian, we have numerically studied the AKLT
model and the Ising PEPS. We found that the Hamiltonian is
local for systems in a gapped phase with local order, whereas a diverging
interaction length of the Hamiltonian is observed when the system
approaches a phase transition, and topological order is reflected in
a Hamiltonian with fully non-local interactions;  thus, the quantum phase
of the bulk can be read off the properties of the boundary
model.

Our holographic mapping between the bulk and the boundary in the PEPS
formalism has further implications.
 In particular, the contraction of PEPS in
numerical simulations requires to approximate the boundary operator by one
with a smaller bond dimension, which can be done efficiently if the boundary
describes the thermal state of a local Hamiltonian, i.e., for non-critical
systems. Also, since renormalization in the PEPS formalism requires to
discard the degrees of freedom in the bond space with the least
weight~\cite{levinwen},
the duality allows to understand real space renormalization in the bulk
as Hamiltonian renormalization on the boundary.

Our techniques can also be applied to systems in higher dimensions, and in
fact to arbitrary graphs, to relate the boundary of a system with its bulk
properties.  The mapping applies to arbitrary regions in the lattice, such
as simply connected (e.g., square) regions used for instance for the
computation of topological entropies. Also, relating the bulk to the
boundary using the PEPS description can be generalized beyond spin systems
by considering fermionic or anyonic PEPS~\cite{fpeps}, as well as
continous PEPS in the case of field theories~\cite{cmps1,cmps2}.
Finally, when studying edge modes, the
one-dimensional system which describes the physical
boundary is given by a Matrix Product Operator acting on the virtual
boundary state, and thus, the relation between bulk properties and the
virtual boundary implies a relation between the properties of the bulk and
its edge modes physics.

\section*{Acknowledgements}

We acknowledge the hospitality of the Kavli Institute for Theoretical Physics (UC Santa Barbara, USA) where this work was initiated.
DP acknowledges support by the French Research Council (Agence Nationale de la Recherche) under grant No.~ANR~2010~BLANC~0406-01 and thanks IDRIS (Orsay, France) and
CALMIP (Toulouse, France) for the use
of NEC-SX8 and Altix SGI supercomputers, respectively.
NS acknowledges support by the Gordon and Betty Moore Foundation through
Caltech's Center for the Physics of Information, the NSF Grant
No.~\mbox{PHY-0803371}, and the ARO Grant No.~W911NF-09-1-0442.
FV acknowledges funding from the SFB projects Vicom and Foqus and the EC projects QUERG and Quevadis. JIC acknowledges the EC project Quevadis, the DFG Forschergruppe 635, and Caixa Manresa.

\end{document}